\documentclass[journal]{IEEEtran}
\usepackage{cite}
\usepackage{array}
\usepackage{fixltx2e}
\usepackage{amssymb,amsfonts,amsmath}
\usepackage{pifont}     % special symbols 
\usepackage{float}

\usepackage{graphicx}
\usepackage{color}
\usepackage{mathtools}
\usepackage{lipsum}
\usepackage{rotating} % to rotate figure
\newcommand{\REVa}[1]{{\color{black}#1}}
\newcommand{\REVb}[1]{{\color{black}#1}}

\graphicspath{{Figures/}}

\makeatletter
\newcommand*{\defeq}{\mathrel{\rlap{%
                     \raisebox{0.3ex}{$\m@th\cdot$}}%
                     \raisebox{-0.3ex}{$\m@th\cdot$}}%
                     =}
\makeatother

\newcommand{\sm}[1]{ \mbox{\tiny \textit{#1}}}

\DeclareMathOperator*{\argmax}{argmax}

\usepackage{hyperref}
\usepackage[capitalise]{cleveref}

\begin{document}
%\title{Bare Demo of IEEEtran.cls\\ for IEEE Journals}
%****************************************************************************************************
\title{Personalized Radiotherapy Design for Glioblastoma: Integrating Mathematical Tumor Models, Multimodal Scans and Bayesian Inference}
%\author{Michael~Shell,~\IEEEmembership{Member,~IEEE,}
%        John~Doe,~\IEEEmembership{Fellow,~OSA,}
%        and~Jane~Doe,~\IEEEmembership{Life~Fellow,~IEEE}% <-this % stops a space
%\thanks{M. Shell was with the Department
%of Electrical and Computer Engineering, Georgia Institute of Technology, Atlanta,

%%-----------------------------------------------------
\author{Jana Lipkov\'{a}$^{1,2,12}$, 
Panagiotis Angelikopoulos$^{3}$,
Stephen Wu$^{4}$,
Esther Alberts$^{2}$,
Benedikt Wiestler$^{2}$,
Christian Diehl$^{2}$,
Christine Preibisch$^{5}$, 
Thomas Pyka$^{6}$,
Stephanie Combs$^{7}$,
Panagiotis Hadjidoukas$^{8,10}$,
Koen Van Leemput$^{9}$,
Petros~Koumoutsakos$^{10}$,
John Lowengrub$^{11}$,
Bjoern Menze$^{1,12}$
%%%------------------------------------------------
\thanks{$^{1}$ Dept. of Informatics, Technical University Munich (TUM) Germany}%
\thanks{$^2$ Dept. of Neuroradiolog, Klinikum Rechts der Isar, TUM, Germany}
\thanks{$^{3}$ D.E. Shaw Research, L.L.C, USA}%
\thanks{$^{4}$ Institute of Statistical Mathematics, Tokyo, Japan}%
\thanks{$^5$ Dept. of Diagnostic and Interventional Neuroradiology \& Neuroimaging Center \& Clinic for Neurology, Klinikum Rechts der Isar, TUM, Germany}
\thanks{$^6$ Dept. of Nuclear Medicine, Klinikum Rechts der Isar, TUM, Germany}
\thanks{$^7$ Dept. of Radiation Oncology, Klinikum Rechts der Isar, TUM \& Institute of Innovative Radiotherapy, Helmholtz Zentrum Munich \& Deutsches Konsortium f\"ur Translationale Krebsforschung, Germany}%
\thanks{$^8$ IBM Research - Zurich, Switzerland. (IBM, the IBM~logo, and ibm.com are trademarks or registered trademarks of International Business Machines Corporation in the United States, other countries, or both. Other product and service names might be trademarks of IBM or other companies.)}%
\thanks{$^9$ Harvard Medical School, Boston, USA \& Dept. of Applied Mathematics and Computer Science, TU Denmark, Denmark.}%
\thanks{$^{10}$ Computational Science and Engineering Lab, ETH Z\"urich, Switzerland}%
\thanks{$^{11}$ Dept. of Mathematics, Biomedical Engineering, Chemical Engineering and Materials Science \& Center for Complex Biological Systems \& Chao Family Comprehensive Cancer Center, UC, Irvine, USA}%
\thanks{$^{12}$ Institute for Advanced Study, TUM, Germany}%
\thanks{The supplementary materials are available at \url{http://ieeexplore.ieee.org}}
}

%\markboth{Journal of \LaTeX\ Class Files,~Vol.~14, No.~8, August~2015}{}%

% For copyringt in single column
%\IEEEoverridecommandlockouts
%\IEEEpubid{\makebox[\columnwidth]{978-1-5386-5541-2/18/\$31.00~\copyright2018 IEEE \hfill} \hspace{\columnsep}\makebox[\columnwidth]{ }}
% For copyright at the whole page
\IEEEoverridecommandlockouts
\IEEEpubid{\begin{minipage}{\textwidth}\ \\[12pt]
\begin{center}
 Copyright (c) 2019 IEEE. Personal use of this material is permitted. However, permission to use this material for any other purposes must be obtained from the IEEE by sending a request to \url{pubs-permissions@ieee.org}.
 \end{center}
\end{minipage}} 
\markboth{}{}


%This work has been submitted to the IEEE for possible publication. Copyright may be transferred without notice, after which this version may no longer be accessible.

\maketitle

% 150-250 words, now (225)
\begin{abstract}
Glioblastoma is a highly invasive brain tumor, whose cells infiltrate surrounding normal brain tissue beyond the lesion outlines visible in the current medical scans. These infiltrative cells are treated mainly by radiotherapy. Existing radiotherapy plans for brain tumors derive from population studies and scarcely account for patient-specific conditions. Here we provide a Bayesian machine learning framework for the rational design of improved, personalized radiotherapy plans using mathematical modeling and patient multimodal medical scans. Our method, for the first time, integrates complementary information from high resolution MRI scans and highly specific FET-PET metabolic maps to infer tumor cell density in glioblastoma patients.  The Bayesian framework quantifies imaging and modeling uncertainties and predicts patient-specific tumor cell density with credible intervals. The proposed methodology relies only on data acquired at a single time point and thus is applicable to standard clinical settings. An initial clinical population study shows that the radiotherapy plans generated from the inferred tumor cell infiltration maps spare more healthy tissue thereby reducing radiation toxicity while yielding comparable accuracy with standard radiotherapy protocols. Moreover, the inferred regions of high tumor cell densities coincide with the tumor radioresistant areas, providing guidance for personalized dose-escalation. The proposed integration of multimodal scans and mathematical modeling provides a robust, non-invasive tool to assist personalized radiotherapy design.
\end{abstract}


% \fi
\maketitle
%
%
%
%===========================================
% 							 INTRODACTION
\IEEEpubidadjcol
\section{Introduction}
\IEEEPARstart{G}{lioblastoma} (GBM) is the most aggressive and most common type of primary brain tumor, with a median survival of only 15 months despite intensive treatment \cite{Stupp:2014}. The standard treatment consists of immediate tumor resection, followed by combined radio- and chemotherapy targeting the residual tumor. All treatment procedures are guided by magnetic resonance imaging (MRI). In contrast to most tumors, GBM infiltrates surrounding tissue, instead of forming a tumor with a well-defined boundary. 
The central tumor, which is visible on medical scans, is commonly resected. However, the distribution of the infiltrating residual tumor cells in the nearby healthy-appearing tissue, which are likely to contribute to tumor recurrence, is not known.
\\
Current radiotherapy (RT) planning handles these uncertainties in a rather rudimentary fashion. Guided by population-level studies, standard-of-care RT plans uniformly irradiate the volume of the visible tumor extended by a uniform margin \cite{Stupp:2014,Stupp:2005,Burnet:2004}, which is referred as the clinical target volume (CTV). However, the extent of this margin varies by few centimeters even across the official RT guidelines \cite{Paulsson:2014}. Moreover, GBM infiltration is anisotropic and thus a uniform margin very likely does not provide an optimal dose distribution. In addition, GBM invasiveness is highly patient-specific, and thus not all patients benefit equally from the same margin, which impairs comparison and advancement of RT protocols.
\\
Despite treatment almost all GBMs recur \cite{Piroth:2012}. Biopsies \cite{Souhami:2004} and post-mortem studies \cite{Halperin:1989} show that tumor cells can invade beyond the CTV, which reduces RT efficiency, and is a possible cause of recurrence. At the same time radioresistance of tumor cells inside the CTV can also reduce RT efficiency.  Radioresistance tends to occur in regions with complex microenviroment and hypoxia \cite{Rockwell:2009}, both of which are commonly encountered in areas of high tumor cellularity. To address tumor radioresistance, several studies have suggested local dose-escalations \cite{Rockwell:2009,Combs:2010,Tsien:2009}. In these approaches, a boosted dose is delivered into a single or multiple co-centered regions defined by adding uniform margins to the tumor outlines visible in MRI scans \cite{Hingorani:2012}. The Radiation Therapy Oncology Group (RTOG) phase-I-trial \cite{Tsien:2009} showed an increase in median survival of 8 months with dose-escalation. However, no benefit in progression-free survival was observed, indicating a complex relationship between true progression of the underlying disease and the tumor extent visible in MRI scans.
\\
Alternatively, positron emission tomography (PET) scans, which map tumor metabolic activities targeted by specific tracers, can be used to identify radioresistant regions. A promising tracer used in GBM imaging is 18F-fluoro-ethyl-tyrosine (FET) \cite{Rieken:2013}, whose uptake values have been shown to be proportional to tumor cell density, although the constant of proportionality is unknown and patient-specific \cite{Stockhammer:2008,Hutterer:2013}. A prospective phase-II-study \cite{Piroth:2012} demonstrated that dose-escalation based on FET-PET enhancement delineates the tumor structure better than uniform margins, thus leading to lower radiation toxicity. Still, FET-PET based dose-escalation did not increase progression-free survival. One possible explanation is that PET enhances mainly the tumor core, which is usually resected, while the PET uptake values in the remaining tumor periphery coincide with the baseline signal from the healthy tissue. This, together with a rather low resolution of PET scans, limits their ability to fully target radioresistant tumor residuals. This is also consistent with our results.
\\
Standard RT plans can be improved by incorporating information from computational tumor models. These models, calibrated against patient medical scans, provide estimates of tumor infiltration that extend the information available in medical images and can guide personalized RT design. Despite extensive  development of tumor growth models  \cite{Alfonso:2017,Swan:2018,Hodgkinson:2018,Cristini:2010,Mohamed:2005,Hogea:2007} and calibration strategies  \cite{Unkelbach:2014,Konukoglu:2010,Harpold:2007,Jackson:2015,Rockne:2015,Le:2015,Hawkins:2018,Menze:2011}, their translation into clinical practice remains very limited. We postulate that there are (at least) three translational weakness: 1) Most model calibrations rely on data not commonly available in clinical practice. For example, in \cite{Konukoglu:2010,Harpold:2007,Jackson:2015,Rockne:2015,Le:2015,Hawkins:2018,Menze:2011} medical scans with visible tumor progression from at least two time points are used for the model calibration. However, for GBM patients only scans acquired at single preoperative time point are available. 2) Models are based on simplified assumptions motivated by \textit{in-vitro} studies. For instance, it is frequently assumed that the tumor cell density is constant along the tumor borders visible on MRI scans (e.g.,\cite{Unkelbach:2014,Konukoglu:2010,Harpold:2007,Jackson:2015,Rockne:2015,Le:2015,Hawkins:2018}). However, the tumor cell density varies significantly along the visible lesion borders due to anatomical restrictions and anisotropic tumor growth. 3) Even if advanced calibration techniques as in \cite{Menze:2011} are used, it is not clear how robust the model predictions are and what benefits they offer over the standard treatment protocols.
\\
Here, we address these translational issues and provide clinically relevant patient-specific tumor predictions to improve personalized RT design. We present a Bayesian machine learning framework to calibrate tumor growth models from multimodal medical scans. We show that an integration of information from complementary structural MRI and functional FET-PET metabolic maps enables the robust inference of the tumor cell densities from scans acquired at single time point. To the best of our knowledge, this is the first study making joint use of FET-PET and MRI scans for the patient-specific calibration of a tumor growth model. Our Bayesian approach infers modeling and imaging parameters under uncertainties arising from measurement and modeling errors. We propagate these uncertainties through the computational tumor model to obtain robust estimates of the tumor cell density together with \REVa{credible} intervals that can be used for personalized RT design. The patient-specific tumor estimates offer an advantage in determining margins of CTV as well as regions for dose-escalation. A clinical study is used to assess benefits of the personalized RT design over standard treatment protocols.
\\
In the remainder of the paper, \cref{sec:Bayes} introduces the Bayesian framework for model calibration, including the tumor growth and imaging models. The results are presented in \cref{sec:Results} where the framework is applied to synthetic and clinical data, followed by a personalized RT study. Conclusions are presented in \cref{sec:Conclusion}. Additional technical details are given in the Supplementary Materials (SM) available in \url{http://ieeexplore.ieee.org}.
%**************************************************************************************************************
%
%
%
%
%
%
%
%
%
%===========================================
% 							 METHOD
%===========================================
\section{Bayesian model calibration}\label{sec:Bayes}
\begin{sidewaysfigure*}
%%~~~~~~~~~~~~~~~~~~~~~~~~~~~~~~~~~~~~~~~~~~~~~~~~~~ 
%\begin{sidewaysfigure*}
    \centering
   \includegraphics[width=1.0\linewidth]{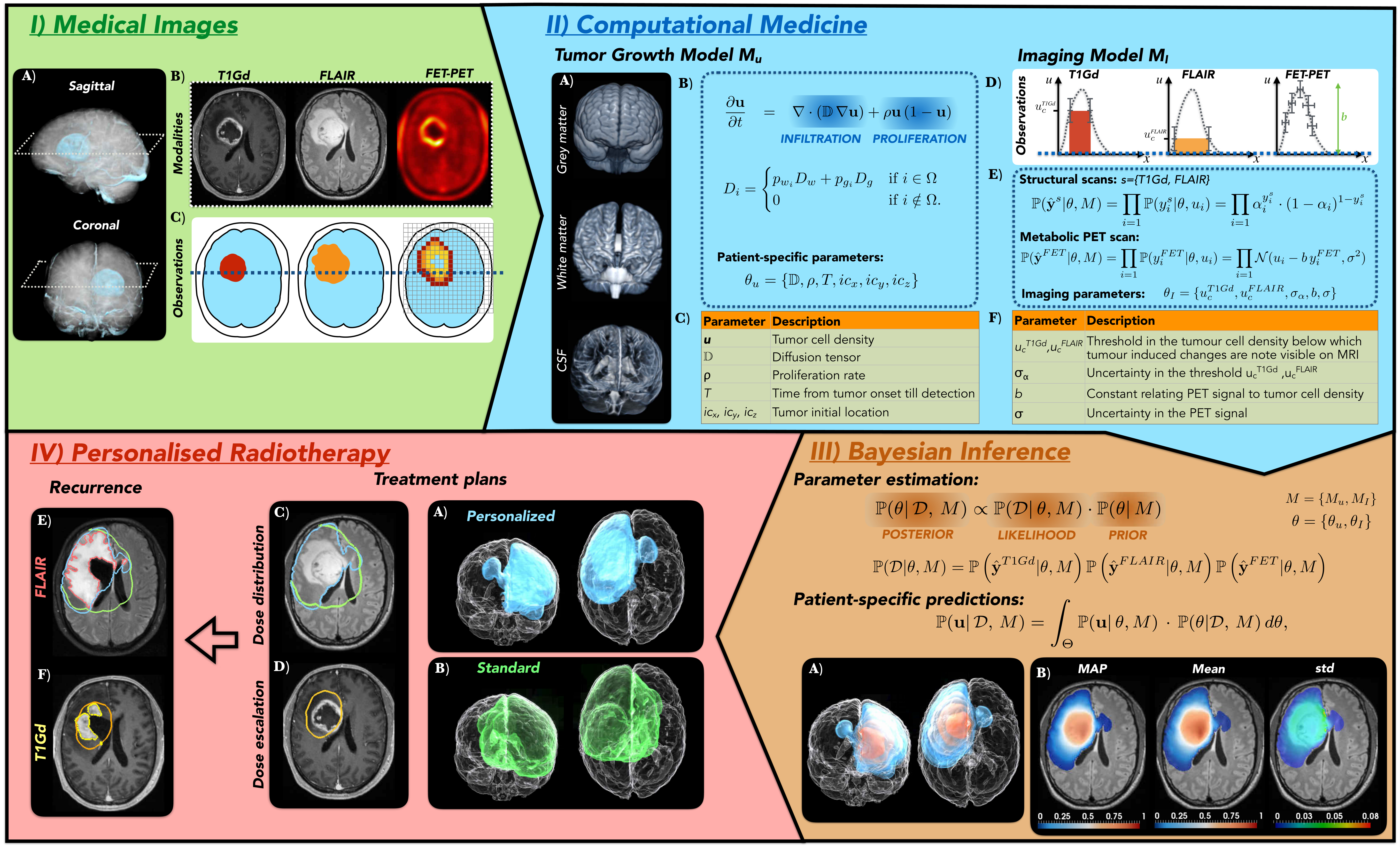}
\caption{Overview of the inference framework. 
\textit{\textbf{I) Medical images}} show preoperative patient scans:  \textit{(I.A)} 3D reconstruction of T1Gd images and \textit{(I.B)} slices across all modalities. Tumor observations (e.g., segmentations) extracted from each modality are illustrated in \textit{(I.C)}. 
\textit{\textbf{II) Computational medicine}} includes a tumor growth model $M_u$ \textit{(II.B)}, which simulates tumor evolution in the patient anatomy \textit{(II.A)}, and imaging model $M_I$ \textit{(II.E)} which relates the image observations with the modeled tumor cell density $\textbf{u}$. Subplot \textit{(II.D)}  shows a schematic representation of the actual tumor cell density along the dashed line shown in \textit{(I.C)} and its relation to each tumor observation available from the medical scans. The unknown, patient-specific parameters for each model are listed in Tables \textit{(II.C,F)}.
\textit{\textbf{III) Bayesian inference}} is used to identify the probability distributions of the unknown parameters, accounting for the modeling and measurement uncertainties. Parametric uncertainties are then propagated to obtain robust predictions about patient-specific tumor cell densities. The most probable tumor cell density, given by the maximum a posterior (MAP) estimate, is shown in \textit{(III.A,B)}, while the mean and standard deviation (std) are shown in \textit{(III.B)}. 
\textit{\textbf{IV) Personalized Radiotherapy}} uses the patient-specific predictions to improve dose-distribution and escalation design. A comparison of a standard and personalized dose distributions is shown in \textit{(IV.A-C)}. The regions of estimated high tumor cell densities are marked by the orange isocontour in \textit{(IV.D,F)}. Observed tumor recurrences in T1Gd and FLAIR, marked by the pink and yellow curves in \textit{(IV.E,F)}, are used to compare the treatment plans. The personalized plan spares more healthy tissue, while achieving tumor coverages comparable to the standard protocol and guides the design of personalized dose escalation plans.}
%\label{fig:Overview}
%\end{sidewaysfigure*}
%~~~~~~~~~~~~~~~~~~~~~~~~~~~~~~~~~~~~~~~~~~~~~~~~~~

\label{fig:Overview}
\end{sidewaysfigure*}

The Bayesian framework we develop combines a deterministic model $M_u$ for tumor growth with a stochastic imaging model $M_{\mathcal{I}}$ relating model predictions with tumor observations available from patient medical scans. Bayes theorem is used to estimate the probability distribution of the unknown parameters of both models, accounting for modeling and measurement uncertainties. Identified parametric uncertainties are propagated to obtain robust patient-specific tumor predictions. An overview of the framework is given in \cref{fig:Overview}.
%******************************************************************************
%
%~~~~~~~~~~~~~~~~~~~~~~~~~~~~~~~~~~~~~~~~~~~~~~~~~~ 
%~~~~~~~~~~~~~~~~~~~~~~~~~~~~~~~~~~~~~~~~~~~~~~~~~~
%
%**************************************************************************************************************
\subsection{Tumor growth model}
Many tumor growth models are based on the Fisher-Kolmogrov (FK) equation \cite{Harpold:2007}, which captures the main tumor behaviour: proliferation and infiltration. We use FK equation to describe the tumor model $M_u$. The equation is solved in a patient-specific brain anatomy reconstructed from MRI scans, where each voxel corresponds to one simulation grid point. Let $\Omega\in\mathbb{R}^3$ be the brain anatomy consisting of white and grey matter and $u_i(t) \in [0,1]$ be normalized tumor cell density at time $t$ and voxel $i$ at location $(i_x,i_y,i_z) \in \Omega$, where $i=\{1,\cdots,N \}$ is index across all voxels. The dynamics of the tumor cell density $\mathbf{u} \defeq \{ u_i(t) \}_{i=1}^{N} $ is modeled as:
%-----------------------------------------------------------------------
\begin{align}
\small
\frac{\partial \mathbf{u}}{\partial t} &= \nabla \cdot \left( \mathbb{D} \, \nabla \mathbf{u} \right) + \rho \mathbf{u} \left( 1 - \mathbf{u} \right) \quad \textit{in } \Omega, \label{eq:FK} \\
\nabla \mathbf{u} \cdot \vec{n} &= 0 \quad\textit{   in }  \partial \Omega . \label{eq:FKBC}
\end{align}
%-----------------------------------------------------------------------
The term $\rho\,[1/day]$ denotes proliferation rate. The tumor infiltration into the surrounding tissues is modeled by the tensor $\mathbb{D} = \{ D_i \,\mathbb{I} \}_{i=1}^{\sm{N}}$ where $\mathbb{I}$ is a $3\times 3$ identity matrix and
%-----------------------------------------------------------------------
\begin{equation}\label{eq:D}
\small
D_{i} =
  \begin{cases}
  p_{w_i} D_w + p_{g_i} D_g 	  & \quad  \text{if } i  \in       \Omega \\
  0           & \quad  \text{if } i  \notin \Omega.  \\
   \end{cases} 
\end{equation}
%-----------------------------------------------------------------------
The terms $p_{w_i}$ and $p_{g_i}$ denote percentage of white and grey matter at voxel $i$, while $D_{w}$ and $D_{g}$ stand for tumor infiltration in the corresponding matter. We assume  $D_w=~10 D_g \,[mm^2/day]$ \cite{Menze:2011}. The skull and ventricles are not infiltrated by the tumor cells and act as a domain boundary with an imposed no-flux boundary condition \cref{eq:FKBC}, where $\vec{n}$ is the outward unit normal to $\partial \Omega$. The tumor is initialized at voxel $(ic_x, ic_{y}, ic_{z})$ and its growth is modeled from time $t=0$ until detection time $t=T \,[day]$. The parameters $\theta_{u} = \{ D_{w}, \rho, T, ic_{x}, ic_{y}, ic_{z} \}$ are considered unknown and patient-specific. The model is implemented in a 3D extension of the multi-resolution adapted grid solver \cite{Rossinelli:2015} with a typical simulation time of 1-3 minutes using 2 cores. An overview of the model $M_u$ and its parameters is shown in \cref{fig:Overview} \textit{(II. A-C)}.
%********************************************************************************************************************
%
%
%
%
%
%
%
%**************************************************************************************************************
\subsection{Multimodal imaging model}\label{subsec:IM}
We consider routinely acquired T1 gadolinium enhanced (T1Gd) and fluid attenuation inversion recovery (FLAIR) MRI scans in combination with FET-PET maps. A stochastic imaging model $M_{\mathcal{I}}$ is designed to relate model predictions of the tumor cell density $\textbf{u}$ and the tumor observations $\mathcal{D}=~\{\hat{\textbf{y}}^{\sm{T1Gd}}, \hat{\textbf{y}}^{\sm{FLAIR}},\hat{\textbf{y}}^{ \sm{FET}} \}$ available from the medical scans. 
\REVa{
Here, $\hat{\textbf{y}}$ denotes a vector of tumor observations obtained from a certain image modality, $y_i$ is an entry in $\hat{\textbf{y}}$ and $i$ enumerates all voxels in the given image. A voxel $i$ corresponds to the same location $(i_x,i_y,i_z)$ in each scan and the simulation domain $\Omega$.} See \cref{fig:Overview} for an overview of all the imaging modalities \textit{(I.B)}, corresponding tumor observations \textit{(I.C)} and their relation to tumor cell density \textbf{u} \textit{(II.D)}.
\\
%**************************************************************************************************************
%
%
%**************************************************************************************************************
The MRI scans provide morphological information about the visible tumor in the form of binary segmentations. The segmentation $\hat{\textbf{y}}^s, s\in\textit{\{T1Gd, FLAIR\}}$ assigns a label $y^{s}_{i}=1$ to each voxel with visible tumor and $y^{s}_{i}=0$ otherwise. The probability of observing a segmentation $\hat{\textbf{y}}^s$ with a simulated tumor cell density $\textbf{u}$ is modeled by a Bernoulli distribution~\cite{Menze:2011}:
%-----------------------------------------------------------------------
\begin{equation}\label{eq:LikelihoodSeg}
\small
\mathbb{P}( \hat{\textbf{y}}^s | \theta, M) = \prod_{i=1}^{N} \mathbb{P}(y_{i}^{s} | \theta, u_{i} ) = \prod_{i=1}^{N} \alpha_{i}^{y_{i}^s} \cdot (1 - \alpha_{i} )^{1-y_{i}^s}.
\end{equation}
%-----------------------------------------------------------------------
Here $\alpha_{i}$ is the probability of observing the tumor in the MRI scan and it is assumed to be a double logistic sigmoid: 
%-----------------------------------------------------------------------
\begin{equation}\label{eq:Sigmoid}
\small
\alpha_{i}(u_{i}, u_{c}^{s}) = 0.5 + 0.5 \cdot \text{sign} (u_{i} - u^{s}_{c} ) \left(  1 - e^{ -\frac{(u_{i} - u^{s}_{c})^2}{\sigma_{\alpha}^{2}} } \right),
\end{equation}
%-----------------------------------------------------------------------
where $u_c^{s}$ denotes an unknown cell density threshold below which tumor cells are not visible in the MRI scan, while the term $\sigma^2_{\alpha}$ represents uncertainty in $u^{s}_{c}$. The parameters $\{ \theta_{\mathcal{I}^{\sm{T1Gd}}}, \theta_{\mathcal{I}^{\sm{FLAIR}}}\} = \{u_{c}^{\sm{T1Gd}}, u_{c}^{\sm{FLAIR}}, \sigma^{2}_{\alpha}\}$ are assumed unknown and patient-specific.\\
%**************************************************************************************************************
%
%
%**************************************************************************************************************
The FET-PET signal is proportional to tumor cell density with an unknown constant of proportionality \cite{Stockhammer:2008,Hutterer:2013}. Let $\hat{\textbf{y}}^{\sm{FET}}$ be the normalized FET-PET signal after subtracting the patient-specific baseline signal from healthy tissue
i.e., $y_{i}^{\sm{FET}}\in [0,1]$ and $b$ the corresponding constant of proportionality. We assume that $y^{\sm{FET}}_i$ can be related with the modeled tumor cell density $u_i$ as
%%-----------------------------------------------------------------------
\begin{equation}
\small
y_i^{\sm{FET}} = \frac{1}{b} u_i + \varepsilon,
\end{equation}
%%-----------------------------------------------------------------------
where $\varepsilon$ is prediction error accounting for modeling and measurement uncertainties. Because of the noisy nature of the PET scan, the error term is assumed to be a normal distribution $\varepsilon\sim\mathcal{N}(0,\sigma^2)$. The probability of observing the PET signal $\hat{\textbf{y}}^{\sm{FET}}$ with the simulated tumor cell density $\textbf{u}$ is then modeled~as
%%-----------------------------------------------------------------------
\begin{equation}\label{eq:LikelihoodPET}
\small{
\mathbb{P}( \hat{\textbf{y}}^{\sm{FET}}| \theta, M)=\prod_{i=1}^{N}\mathbb{P}(y_{i}^{\sm{FET}}| \theta, u_{i} )=\prod_{i=1}^{N } \mathcal{N}\left(y^{\sm{FET}}_{i} - \frac{1}{b} u_i, \, \sigma^{2}\right).
}
\end{equation}
%-----------------------------------------------------------------------
The PET scan, acquired at $4\,mm$ resolution, is registered to the MRI scans with $1\,mm$ resolution. To justify the product in \cref{eq:LikelihoodPET} only voxels separated by distance $4\,mm$ are used. The parameters $\theta_{\mathcal{I}^{\sm{FET}}}= \{b,\sigma\}$ are unknown and patient-specific. An overview of the imaging model $M_{\mathcal{I}}$ and its parameters is shown \cref{fig:Overview} \textit{(II. D-F)}.
%**********************************************************************************************************
%
%
%
%
%**********************************************************************************************************
\subsection{Parameters estimation and uncertainty propagation}\label{subsec:UQ+P}
The parameters $\theta=\{\theta_u,\theta_{\mathcal{I}}\}$ of the model $M=\{M_u,M_{\mathcal{I}}\}$ where $\mathcal{I}=\{ \mathcal{I}^{\sm{T1Gd}},\mathcal{I}^{\sm{FLAIR}},\mathcal{I}^{\sm{PET}} \}$, are assumed unknown and a probability distribution function (PDF) is used to quantify their plausible values. A \textit{prior} PDF $\mathbb{P}(\theta | M)$ is used to incorporate any prior information about $\theta$. Bayesian model calibration updates this prior information based on the available data $\mathcal{D}$. The updated \textit{posterior} PDF is computed by the Bayes theorem:
%-----------------------------------------------------------------------
\begin{equation}\label{eq:Bayes}
\small
\mathbb{P}(\theta|\,\mathcal{D},\,M) \propto \mathbb{P}(\mathcal{D}| \,\theta, M) \cdot \mathbb{P}(\theta |\, M),
\end{equation}
%-----------------------------------------------------------------------
where $\mathbb{P}(\mathcal{D}|\theta,M)$ is the \textit{likelihood} of observing data $\mathcal{D}$ from the model $M$ for a given value of $\theta$. Since each of the medical scans captures a different physiological process, the tumor observations are assumed independent and the likelihood function can be expressed as:
%-----------------------------------------------------------------------
\begin{equation}\label{eq:Likelihood}
\small
{\mathbb{P}(\mathcal{D}|\theta,M) =  \mathbb{P}\left(  \hat{\textbf{y}}^{\sm{T1Gd}} | \theta, M \right) \cdot \mathbb{P}\left(  \hat{\textbf{y}}^{\sm{FLAIR}}| \theta, M \right) \cdot \mathbb{P}\left( \hat{\textbf{y}}^{\sm{FET}}| \theta, M\right).} \nonumber 
\end{equation}
%-----------------------------------------------------------------------
The prior PDF is assumed uniform with details specified in the SM. Since an analytical expression for \cref{eq:Bayes} is not available, sampling algorithms are used to obtain samples $\theta^{(l)},$ $ l\in\{1,\cdots,S\}$ from the posterior $\mathbb{P}(\theta | \mathcal{D},M)$. We use Transitional Markov Chain Monte Carlo (TMCMC) algorithm \cite{Ching:2007} which iteratively constructs series of intermediate PDFs: 
%--------------------------------------------------------
\begin{equation}
\small
\mathbb{P}_{j}(\theta | \mathcal{D}, M) \sim  \mathbb{P}( \mathcal{D}|\theta, M)^{p_{j}}  \cdot \mathbb{P}(\theta |M),
\end{equation}
%--------------------------------------------------------
where $ 0=p_{0}<p_{1}<\cdots<p_{m}=1$ and $j=\{1,\cdots,m\}$ is a generation index. The term $p_{j}$ controls the convergence of the sampling procedure and is computed automatically by the TMCMC algorithm. TMCMC method constructs a large number of independent chains that explore parameter space more efficiently than traditional sampling methods \cite{Ching:2007} and allow parallel execution. We use a highly parallel implementation of the TMCMC algorithm provided by the $\Pi$4U framework \cite{Hadjidoukas:2015}.
\\
%********************************************************************************************************************
%
%
%********************************************************************************************************************
The inferred parametric uncertainties are propagated through the model $M$ to obtain robust predictions about $\textbf{u}$ given by:
%-----------------------------------------------------------------------
\begin{equation}\label{eq:RobustPosterior}
\small
\mathbb{P}(\textbf{u} | \,\mathcal{D}, \, M) =  \int_{\Theta }  \mathbb{P}(\textbf{u} |\,\theta, M)\, \cdot \, \mathbb{P}(\theta | \mathcal{D},\, M)  \, d\theta,
\end{equation}
%-----------------------------------------------------------------------
or by simplified measures such as the mean  $\mu_{\textbf{u}} = E[\textbf{u}(\theta)] \equiv m_{1} $ and variance $\sigma^2_{\textbf{u}} = E[\textbf{u}^{2}(\theta)] - m_{1}^2 \equiv m_2 - m_{1}^2$ derived from the first two moments $m_{k}$, $k=1,2$: 
%-----------------------------------------------------------------------
\begin{equation}\label{eq:Moments}
\small
m_{k}=  \int_{\Theta} \big( \textbf{u}(\theta |M ) \big)^k \cdot  \mathbb{P}(\theta | \mathcal{D},M) \, d\theta \approx  \frac{1}{S} \sum_{l=1}^{S} \left(\textbf{u}(\theta^{(l)} | M) \right)^k,
\nonumber
\end{equation}
%-----------------------------------------------------------------------
where $\Theta$ is the space of all unknown parameters. The most probable tumor cell density estimate, is given by the maximum a posteriori (MAP) defined as $\textbf{u}^{\sm{MAP}} = \argmax_{\theta} \mathbb{P}(\textbf{u} | \,\mathcal{D}, \, M)$.
%**********************************************************************************************************
%
%
%
%
%
%===========================================
% 							 RESULTS
%===========================================
\section{Results}\label{sec:Results}
The Bayesian framework described in the previous section is first applied to synthetic data to test sensitivity of the inference and to show the role of multimodal image information on the model calibration. Afterwards, clinical data are used to infer patient-specific tumor cell densities and to design personalized RT plans. Tumor recurrence patterns are used to compare the proposed and standard RT plans. The software and data used in this paper are publicly available\footnote{\url{https://github.com/JanaLipkova/GliomaSolver}}.
%****************************************************************************************************************************************************************************
%
%
%
%
%
%
%%================   	Table  	========================
\begin{table*}
\small
\centering 
\caption{Results of the Bayesian calibration for the synthetic case generated with the ground truth (GT) values. Reported is maximum a posteriori (MAP), mean and standard deviation (std). The units are $D_w\sim mm^2/day$; $\rho\sim 1/day$; $T\sim day$; and $ic_x$, $ic_y$, $ic_z\sim mm$.
%The space units are in $mm$, while time in $days$.
}
\begin{tabular*}{\hsize}{@{\extracolsep{\fill}} lccccccccccccc }
\hline 
\hline 
   & $D_{w}$ & $\rho$ & $T$ & $ic_{x}$ & $ic_{y}$ &  $ic_{z}$ & $\sigma$ & $b$ &$u_{c}^{\textit{T1Gd}}$ & $u_c^{\textit{FLAIR}}$ & $\sigma^{2}_{\alpha}$ 
\cr 
\hline 
%\hline 
GT    &  0.1300 &  0.0250 &  302.00 &  80.64 &  171.52 &  128.00 &  0.0230 &  0.8800 &  0.7000 &  0.2500 &  - &  \cr 
MAP   &  0.1226 &  0.0296 &  271.13 &  80.61 &  171.55 &  127.92 &  0.0280 &  0.9884 &  0.8440 &  0.4874 &  0.0502 &  \cr 
mean  &  0.1543 &  0.0361 &  253.85 &  80.64 &  171.47 &  127.80 &  0.0294 &  0.9775 &  0.8325 &  0.4841 &  0.0505 &  \cr 
std   &  0.0530 &  0.0125 &  107.59 &  0.1280 &  0.1024 &  0.2048 &  0.0022 &  0.0139 &  0.0132 &  0.0076 &  0.0004 &  \cr 
\hline 
\hline 
\end{tabular*} 

\label{table:SynResults}
\end{table*} 
%%================================================
%
%
%~~~~~~~~~~~~~~~~~~~~~~~~~~~~~~~~~~~~~~~~~~~~~~~~~~ 
\begin{figure}
\centering
\includegraphics[width=1\linewidth]{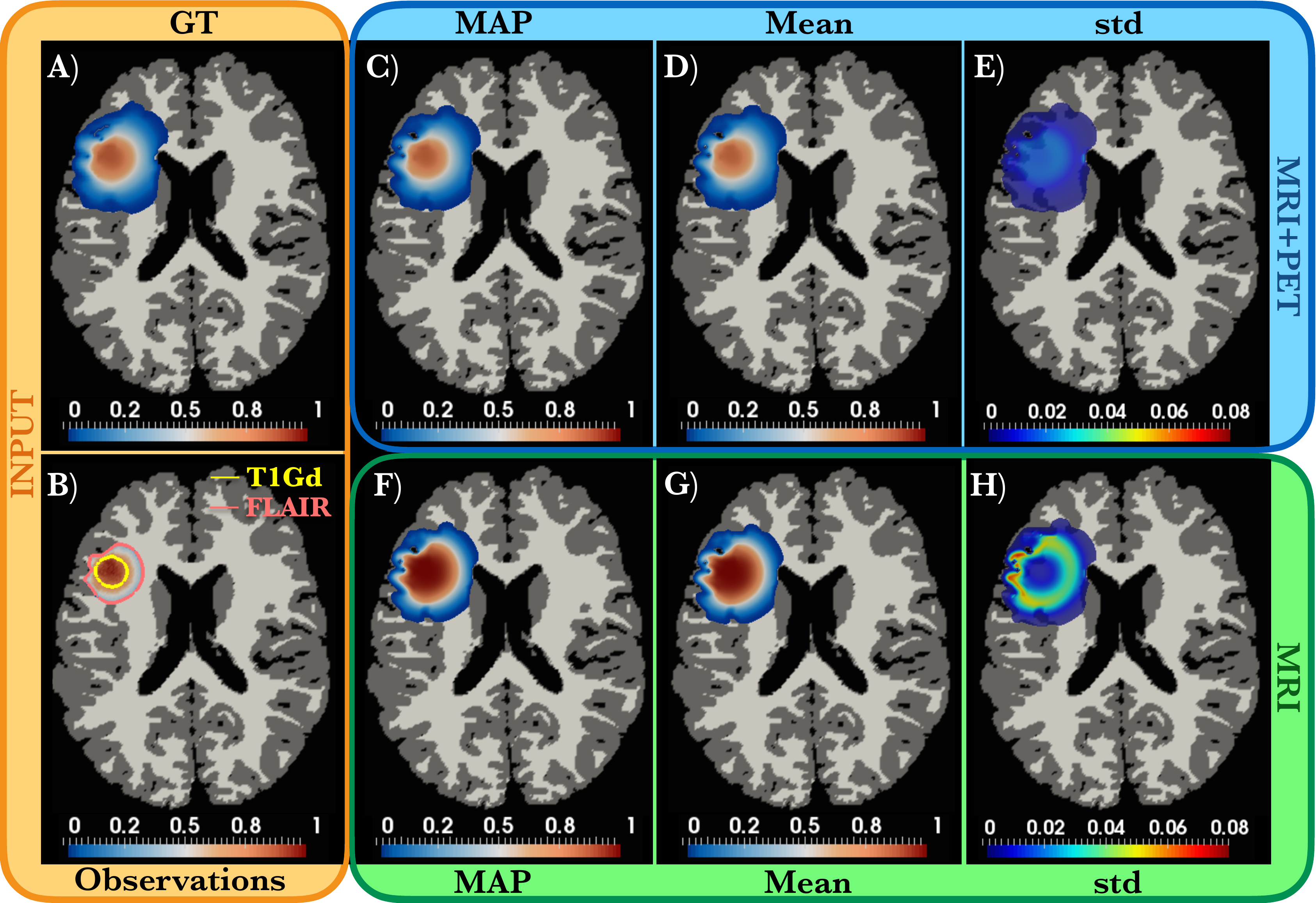}
\caption{Synthetic test case. \textbf{Orange box:} A 2D slice of the synthetic ground truth (GT) tumor cell density \textit{(A)} and corresponding image observations \textit{(B)}: the normalized FET-PET signal with additive noise (red-blue color scale) and the outlines of the T1Gd (yellow) and FLAIR (pink) binary tumor segmentations. \textbf{Blue box:} Results of the Bayesian calibration with multimodal data. The results are in close agreement with GT data. \textbf{Green box:} Calibration results using only the MRI data, which do not provide enough information to recover the tumor cell density profile correctly.}
\label{fig:SynResults}
\end{figure}
%~~~~~~~~~~~~~~~~~~~~~~~~~~~~~~~~~~~~~~~~~~~~~~~~~~
%
%
%~~~~~~~~~~~~~~~~~~~~~~~~~~~~~~~~~~~~~~~~~~~~~~~~~~ 
\begin{figure}%keepaspectratio
\centering
\includegraphics[width=1.00\linewidth]{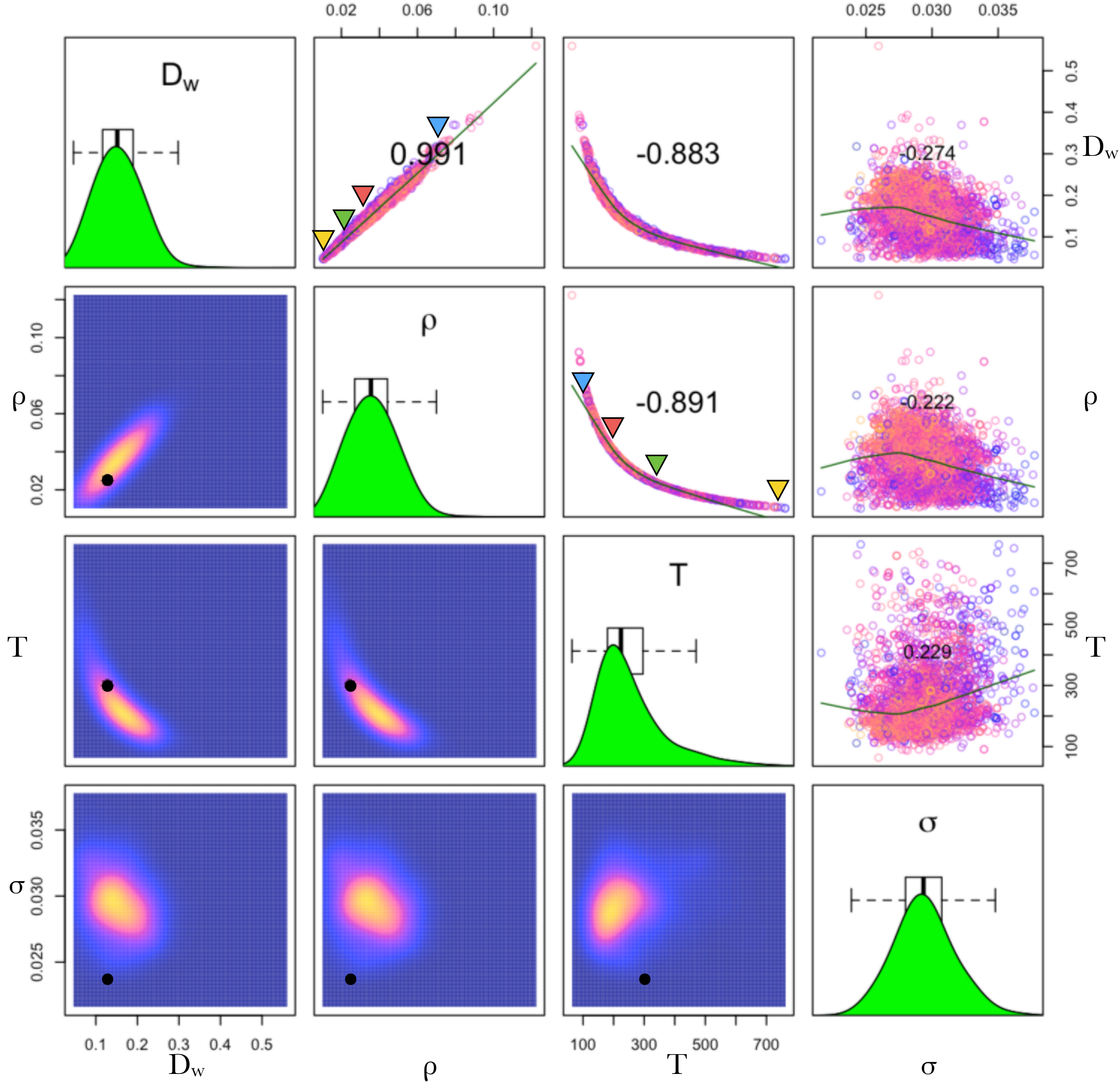}
\caption{The results of the Bayesian calibration for the synthetic case. \textbf{Above the diagonal:} Projection of the TMCMC samples of the posterior distribution $\mathbb{P}(\theta |\mathcal{D},M)$ in 2D space of the indicated parameters. The colors indicate likelihood values of the samples. The number in each plot shows the Pearson correlation coefficient between the parameter pairs. The colored triangles mark the four selected parameters used in \cref{fig:vel}. \textbf{Diagonal:} Marginal distributions obtained with Gaussian kernel estimates. Boxplot whiskers demarcate the 95\% percentiles. \textbf{Below the diagonal:} Projected densities in 2D parameter space constructed by 2D Gaussian kernel estimates. The black dots mark the values used to generate the synthetic data. }
\label{fig:S01samples}
\end{figure}
%~~~~~~~~~~~~~~~~~~~~~~~~~~~~~~~~~~~~~~~~~~~~~~~~~~
%
%
%
%~~~~~~~~~~~~~~~~~~~~~~~~~~~~~~~~~~~~~~~~~~~~~~~~~~ 
\begin{figure}%keepaspectratio
\centering
\includegraphics[width=1\linewidth]{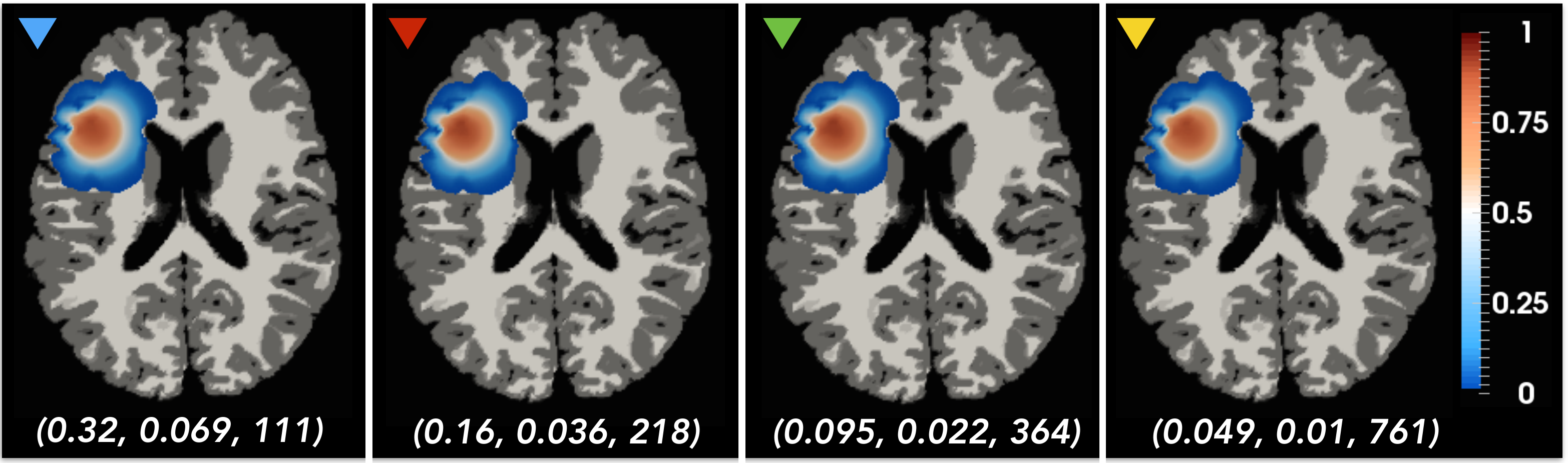}
\caption{Insensitivity of the tumor cell density to the speed of the growth. Shown are slices of the tumor cell densities computed with different combinations of parameters $(D_w,\rho,T)$ as listed at the bottom of each plot. These correspond to the colored triangles in \cref{fig:S01samples}. Despite significant variation in the parameter values, all combinations lead to very similarly-appearing tumors. In the absence of temporal information, \REVa{the time dependent parameters are not identifiable, since} the model calibration cannot distinguish between compensating effects among the parameters that affect the dynamics. As shown here, tumors with similar $D_w/\rho$ and $T\rho$ values appear very similar to one another (here $D_w/\rho\approx 4.5$, $T\rho\approx 7.7$). Hence, the Bayesian calibration identifies the probability distribution of all the plausible values.}
\label{fig:vel}
\end{figure}
%~~~~~~~~~~~~~~~~~~~~~~~~~~~~~~~~~~~~~~~~~~~~~~~~~~
%
%
%
%
%
%****************************************************************************************************************************************************************************
\subsection{Sensitivity study}\label{subsec:SS}
The model $M_{u}$ is used to generate a 3D synthetic tumor in a brain anatomy obtained from \cite{Cocosco:1997} using the parameters reported in \cref{table:SynResults}. A 2D slice of the simulated 'ground-truth' (GT) tumor cell density is shown in \cref{fig:SynResults} \textit{(A)}. The synthetic T1Gd and FLAIR tumor segmentations are constructed by thresholding the GT tumor cell density at $u_{c}^{ \sm{T1Gd}}= 0.7$ and $u_{c}^{\sm{FLAIR}}=0.25$. The FET-PET signal is designed by taking the GT tumor cell density within the T1Gd and FLAIR segmentations, adding Gaussian noise with zero mean and standard deviation (std) $\sigma$, and normalizing the result. The value of $\sigma$ is chosen as average std of the FET signal from the healthy brain tissue. The generated synthetic image observations are shown in \cref{fig:SynResults} \textit{(B)}.
\\
A sensitivity study for the number of samples is performed, indicating that 6000 samples is adequate for the model. The manifold of the inferred probability distribution is presented in \cref{fig:S01samples} and the calibrated parameters are given in \cref{table:SynResults}. As seen from the probability distribution manifold, tumor observations from a single time point do not contain enough information to infer time dependent parameters $(D_{w},\rho,T)$ exactly, since different combinations of these parameters can generate the same tumor cell density as shown in \cref{fig:vel}. \REVa{The lack of identifiability of $(D_{w},\rho,T)$} poses a challenge for calibration approaches searching only for a single value of $\theta$. Instead, Bayesian calibration provides fairer estimate; the inferred probability distribution shows a strong correlation between the parameters $(D_{w},\rho,T)$, while the high std values imply low confidence in these parameters. On the other hand, parameters that affect the tumor spatial pattern, e.g. $(ic_x,ic_y,ic_z)$, are identified with high accuracy, which is reflected by their low std. The image related parameters $(u_{c}^{\sm{T1Gd}}, u_c^{\sm{FLAIR}}, b, \sigma)$ are slightly overestimated due to the assumed correlation length and the effect of the complex brain anatomy. The role of the anatomy is discussed further in the SM.
\\
The inferred parametric uncertainties are propagated to obtain robust posterior predictions about the tumor cell density shown in \cref{fig:SynResults} \textit{(C-E)}. Despite the large parametric uncertainties, the MAP and mean tumor cell density estimates are almost indistinguishable from the GT tumor. The low std values imply that, using our Bayesian formulation, the information contained in multimodal data is sufficient to infer tumor cell density from single time point scans.
\\
For comparison, if the model calibration is performed only with the MRI data, i.e. $\mathcal{D} = \{ \hat{\textbf{y}}^{\sm{T1Gd}},  \hat{\textbf{y}}^{\sm{FLAIR}} \} $, the estimated tumor cell densities shown in \cref{fig:SynResults} \textit{(F-G)} deviate from the GT tumor mainly in the central part of the lesion, which is also consistent with the regions of high std shown in \cref{fig:SynResults} \textit{(H)}. Nonetheless, the outlines of the predicted tumor are similar to those of the GT tumor. This is because the tumor morphology is mainly constrained by the MRI data, since the FET-PET signal coincides with the baseline signal of the healthy tissue in the regions of lower tumor infiltration. On the other hand, the FET-PET signal constrains the tumor cell density profile in the regions of high tumor infiltration. This highlights the importance of integrating structural and functional image information for the model calibration when dealing with single time point data.
%************************************************************************************************************
%
%
%
%
%
%
%~~~~~~~~~~~~~~~~~~~~~~~~~~~~~~~~~~~~~~~~~~~~~~~~~~ 
\begin{figure*}%[tbhp]
\centering
\input{Figures/PatientInferenceFigure.tex}
\label{fig:PI}
\end{figure*}
%~~~~~~~~~~~~~~~~~~~~~~~~~~~~~~~~~~~~~~~~~~~~~~~~~~
%
%
%
%************************************************************************************************************
\subsection{Patient study}\label{subsec:PS}
A retrospective clinical study is conducted on 8 patients diagnosed with GBM. Scans of the patients \textit{P1-P8} are shown \cref{fig:PI} and the details about acquisition protocols \REVa{and image processing} are reported in the SM. All patients received the standard treatment, surgery followed by combined radio- and chemotherapy \cite{Stupp:2005}. There was no visible tumor after the treatment and patients were regularly monitored for recurrence. The preoperative scans shown in \cref{fig:PI} \textit{(A-D)} are used for the Bayesian inference. The calibrated parameters are reported in Table 1 in the SM and the posterior, patient-specific predictions for the tumor cell densities are shown in \cref{fig:PI} \textit{(E-G)}. These patient-specific predictions provide estimates about the possible tumor cell migration pathways in the surrounding of the visible tumor, constrained by the patient anatomy and the available tumor observations. The predicted tumor infiltration pathways can be validated by the patterns of the first detected tumor recurrence shown in \cref{fig:PI} \textit{(H)}, where the outlines of the predicted infiltrations (blue) and recurrence tumors (pink) are depicted. For patients \textit{P5,P7,P8} the model accurately predicts tumor infiltration also inside the healthy-appearing collateral hemisphere, whereas for cases \textit{P1-P4} the tumor predictions are correctly restricted only to one hemisphere. Moreover, despite a similar appearance of the preoperative tumors in patients \textit{P1} and \textit{P2}, the model correctly predicts more infiltrative behaviour for the patient \textit{P1}, which is consistent with the recurrence pattern.  The high confidence in the predictions is reflected by low std shown in \cref{fig:PI} \textit{(F)}.
%************************************************************************************************************
%
%
%
%
%
%************************************************************************************************************
\subsection{Personalized radiotherapy design}\label{subsec:RT}
The patient-specific tumor cell density predictions can be used to design margins of the CTV and to identify high cellularity regions that could mark areas of increased radioresistance. The personalized RT plan can be based either on \textit{the most probable scenario} given by MAP estimate or \textit{the worst case scenario} given as a sum of the mean and std of the tumor cell density. Since in the present study, the mean and MAP estimates are very similar, and the std values are small, the MAP estimates are used. An overview of the proposed personalized RT design is shown in \cref{fig:Overview} \textit{(IV)}, while the details are described in the following subsections. The tumors recurrence patterns are used to assess the benefits of the proposed RT plan over the standard treatment protocol. For evaluation purposes, all recurrence scans are registered to the preoperative anatomy. To prevent registration errors arising from mapping the anatomy with the resection cavity to the preoperative brain, rigid registration is used. This provides a sufficient mapping for most cases, however it cannot capture the post-treatment tissue displacement around the ventricles in patients \textit{P7-P8}, making the mapping less accurate in these regions. The design of methods that provide robust registration between pre- and post-operative brain anatomies is still an open problem.
\\
%************************************************************************************************************
%
%
%
%
%
%************************************************************************************************************
\subsubsection{Dose distribution}
\begin{figure}[t]%[tbhp]
\centering
\includegraphics[width=0.80\linewidth]{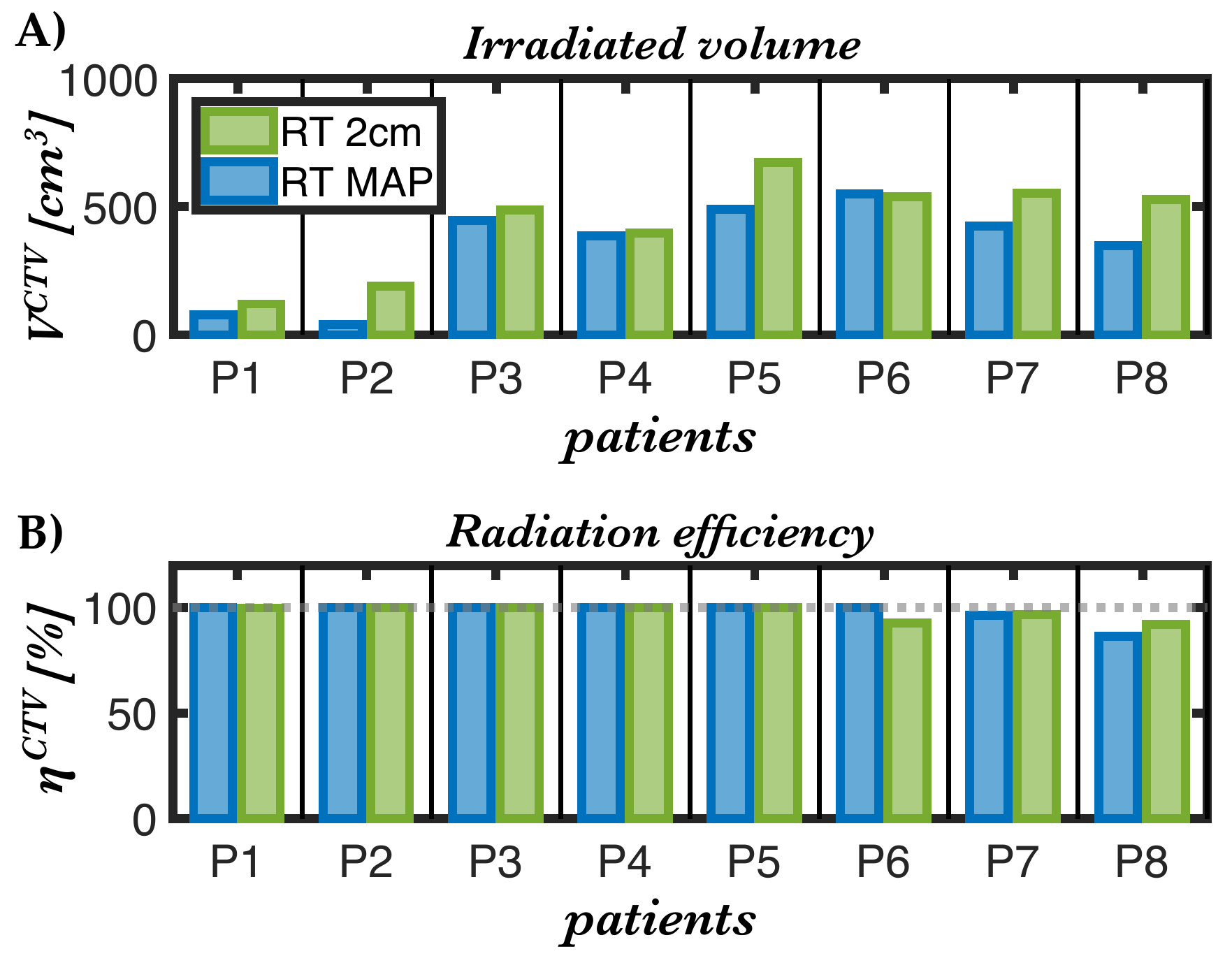}
\caption{A comparison of the RT plan based on the RTOG protocol (green) and MAP estimates (blue). \textit{(A)} The overall irradiated volume $V^{\sm{CTV}}$ and \textit{(B)} the corresponding efficiency $\eta^{\sm{CTV}}$. The $CTV^{\sm{MAP}}$ uses a smaller irradiation volume while having a comparable efficiency as the $CTV^{\sm{RTOG}}$.}
\label{fig:RTerror}
\end{figure}

An ideal CTV covers all the residual tumor, including infiltrating tumor cells that are invisible on the pretreatment imaging scans, while sparing healthy tissue. We use the tumor recurrence pattern to evaluate the efficiency ($\eta^{CTV}$) of the CTV, defined here as the relative volume of the recurrence tumor ($V^{\sm{REC}}$) contained within the CTV:
%-----------------------------------------------------------------------
\begin{equation}\label{eq:efficiency}
\small
\eta^{\sm{CTV}} = \frac{|V^{\sm{REC}} \cap CTV|}{V^{\sm{REC}}} \times 100 \%.
\end{equation}
%-----------------------------------------------------------------------
\Cref{fig:PI} \textit{(H)} shows the FLAIR scans with the first detected tumor recurrence outlined by the pink lines. The margin of the administered $CTV^{\sm{RTOG}}$, designed by the standard RTOG protocol with a $2\,cm$ margin around the visible tumor, is marked by the green lines in \textit{(E,F,H)}. The personalized CTV, referred as $CTV^{\sm{MAP}}$, is constructed by thresholding the MAP tumor cell density at $u=0.1\%$ for all patients. (This value was chosen so that the efficiency of the $CTV^{\sm{MAP}}$ is comparable to that of the $CTV^{\sm{RTOG}}$ ). The outlines of the proposed $CTV^{\sm{MAP}}$ are shown as the blue isocontours in  \cref{fig:PI} \textit{(H-I)}. A visual comparison of the CTVs shown in \cref{fig:PI} \textit{(H)} and \cref{fig:Overview} \textit{(IV)}, and a quantitative comparison presented in \cref{fig:RTerror}, show that the proposed personalized plans spare more healthy tissue, hence reducing radiation toxicity, while maintaining the efficiency of the standard RTOG protocol. Both plans show reduced efficiency for patients \textit{P7-P8}, mainly around the ventricles, which may be caused by misalignment between the preoperative and recurrence anatomies.
\\
These preliminary results imply that the regions predicted as a tumor-free by the model, remain tumor-free and thus the model predictions have potential to guide personalized CTV design. The standard or hospital-specific protocols can be updated by the model predictions to spare brain tissue not infiltrated by the tumor. This can lead to significant savings in the healthy tissue, especially in the cases of large lesions or lesions close to hemispheres separation and other anatomical constraints.
\\
%************************************************************************************************************
%
%
%~~~~~~~~~~~~~~~~~~~~~~~~~~~~~~~~~~~~~~~~~~~~~~~~~~ 
%~~~~~~~~~~~~~~~~~~~~~~~~~~~~~~~~~~~~~~~~~~~~~~~~~~~
%
%
%************************************************************************************************************
\subsubsection{Dose-escalation}
No dose-escalation plan for GBM patients has been yet approved by phase-III-clinical trials. Here, we present a theoretical comparison of two escalation plans targeting high tumor cellularity regions identified by: 1) FET-PET enhancement as proposed in \cite{Piroth:2012} and 2) MAP estimates. We evaluate the efficiency of an escalation plan by its capability of targeting T1Gd-enhanced tumor recurrence regions. In these regions, the recurrent tumor has high cellularity, despite having received the full radiation dose, suggesting tumor radioresistance. \Cref{fig:PI} \textit{(I)} shows the T1Gd scans with the first detected tumor recurrence. The margins of the T1Gd-enhanced tumor recurrence are marked by the yellow lines, while the outlines of the dose-escalation plans designed by the FET-PET enhancements are shown in magenta. The FET enhancements do not fully cover the T1Gd recurrent tumor in patients \textit{P4-P7}, providing a possible explanation for why improvements in progression-free survival have not been observed in \cite{Piroth:2012}. In comparison, the MAP estimates, calibrated by the FET-PET signal, extend the information about the tumor cell density in the periphery of the visible lesion. \Cref{fig:PI} shows two possible dose-escalation plans based on the inferred MAP tumor cell density: \textit{(I)} a single-level dose-escalation based on the thresholded MAP solution with threshold $u=~30\%$ marked by the orange lines and \textit{(J)} a cascaded four-level escalation plan constructed by thresholding the MAP tumor cell density at $u = [0.1, 25, 50, 75] \%$. The optimal design of a personalized dose-escalation plan would require more extensive studies. However, these preliminary results show that the inferred high cellularity regions coincide with the areas of tumor recurrence better than those suggested by the FET-PET enhancement alone. The Bayesian inference framework developed here thus provide a promising tool for a rational dose-escalation design.
%************************************************************************************************************
%
%
%
%
%
%****************************************************************************************************************************************************************************
\section{Conclusion}\label{sec:Conclusion}
We have demonstrated that patient-specific, data-driven modeling can extend the capabilities of personalized RT design for infiltrative brain lesions. We combined patient structural and metabolic scans from a single time point with a computational tumor growth model through a Bayesian inference framework and predicted the tumor distribution beyond the outlines visible in medical scans. The patient-specific tumor estimates can be used to design personalized RT plans, targeting shortcomings of standard RT protocols. The software and data used in this work are publicly released\footnote{\url{https://github.com/JanaLipkova/GliomaSolver}} to facilitate translation to clinical practice and to encourage future improvements. \REVb{In the future, the Bayesian framework developed here could also be extended to predict individual patient responses to RT by incorporating data obtained during the course of treatment as done in \cite{Tariq:2016}, in which non-spatial tumor models were used. In this way, the treatment can be further improved by adaptively refining the RT plans based on the predicted patient responses.} Moreover, the basic FK tumor growth model could be replaced by \REVa{a Fokker-Planck diffusion model \cite{Swan:2018}, which would not increase the number of unknown parameters or affect the computational complexity significantly, but might provide a better description of biological diffusion.} Future work could also incorporate more advanced models, such as \cite{Yan:2017}, that account for cancer stem cells, their progeny and nonlinear coupling between the tumor and the neovascular network. However, it remains to be seen whether scans acquired at a single time point would provide enough information to calibrate the advanced models sufficiently. If not, then simpler, well-calibrated models may prove to be more informative.
Finally, in future studies, the computational framework developed here will be tested on a larger patient cohort and prospective clinical trials will be performed. In summary, the results presented here provide a proof-of-concept that multimodal Bayesian model calibration holds a great promise to assist the development of personalized RT protocols.
%%****************************************************************************************************************************************************************************
%
%
%
%****************************************************************************************************************************************************************************
%            Acknowledgments
%%****************************************************************************************************************************************************************************
\section*{ACKNOWLEDGMENT}
JL acknowledges partial funding from the National Science Foundation-Division of Mathematical Sciences (NSF-DMS)
through grant DMS-1714973 and the Center for Multiscale Cell Fate Research at UC Irvine, which is supported by NSF-DMS (DMS-1763272)
 and the Simons Foundation (594598, QN).
JL additionally acknowledges partial funding from the National Institutes of Health (NIH) through grant
1U54CA217378-01A1 for a National Center in Cancer Systems Biology at the University of California, Irvine, and
NIH grant P30CA062203 for the Chao Comprehensive Cancer Center at the University of California, Irvine.
\\
JL acknowledges the partial funding from the Bavaria California Technology Center (BaCaTec) through grant 6090142.

%****************************************************************************************************************************************************************************
%            Bibliography
%****************************************************************************************************************************************************************************
%\pnasbreak splits and balances the columns before the references.
% If you see unexpected formatting errors, try commenting out this line
% as it can run into problems with floats and footnotes on the final page.
%\pnasbreak
%\bibliographystyle{pnas-new.bst}
\bibliographystyle{IEEEtran}
\bibliography{BibTex/GliomaBibFile}



%\bibliographystyle{IEEEtran}
%\begin{thebibliography}{1}
%\end{thebibliography}

%\vfill

\end{document}

\maketitle

%****************************************************************************************************************************************************************************
%     Subsection: Prior
%****************************************************************************************************************************************************************************
\subsection{Prior PDF}\label{Prior}
The parameters $\theta = \{ \theta_{u}, \theta_{\mathcal{F}^{\sm{T1Gd}}}, \theta_{\mathcal{F}^{\sm{FLAIR}}}, \theta_{\mathcal{F}^{\sm{FET}}} \}$ are assumed independent with uniform prior distribution: $\mathbb{P} \left( \theta | M \right) =  \prod_{j=1}^{11} \mathbb{P}(\theta_{j} | M) $, where
%-----------------------------------------------------------------------
\begin{equation}\label{eq:Prior}
\small
  \mathbb{P}(\theta_{j} | M) = \frac{1}{ \log(\theta^{max}_{j}) - \log(\theta^{min}_{j})}  \quad \textit{for } \theta_{j}^{min} \leq \theta_{j} \leq \theta_{j}^{max},
\end{equation}
%----------------------------------------------------------------------
and zero otherwise. The logarithm ensures a similar order of magnitude for the prior ranges of all considered parameters, which facilitates the sampling procedure. The prior ranges are chosen as follows: $ D_{w} \in \left[ 0.013,  3.8 \right] mm^2/day$, $\rho \in \left[ 0.0027,  0.19 \right] /day$ as reported in \cite{Harpold:2007}, $T \in \left[ 30, T^{max} + 365 \right] \,day$, where $T^{max}$ is the maximum time that a propagation wave with velocity $v^{min}=2\sqrt{min(D_w \rho)} $ needs to travel a distance equal to the maximum radius of the visible tumor  $r^{max}$. A prior for the initial location is taken as a cube with a side $2\,r^{max}$, centered at the tumor center of mass. The imaging model parameters are assumed to be $u_{c}^{T1Gd} \in [0.5,0.85]$, $u^{FLAIR}_{c} \in [0.05, 0.5]$, $\sigma^{2}_{\alpha} \in [0.05, 0.1] $, $\sigma \in [ 0.015, 0.2 ]$ and $b \in [0.5, 1.02]$. The lower bound for the scaling factor $b$ comes from the fact the T1Gd and FET-PET scans usually do not enhance low grade gliomas, suggesting the same lower bound for the both modalities.

\subsection{Bayesian calibration} \cref{table:PatientResults} reports the results of the Bayesian calibration for the clinical data shown in Fig.~5 in the main text.

%%%================   	Table  	========================
\begin{table*} [tbhp!]
\centering 
\caption{
Results of the Bayesian calibration for the patient data shown in Fig.~5 in the main text. The maximum a posterior (MAP) estimates, means and standard deviations (std) of the marginal distribution are shown for all the parameters. The units are $D_w\sim mm^2/day$; $\rho\sim 1/day$; $T\sim day$; and $ic_x$, $ic_y$, $ic_z\sim mm$.
}
\begin{tabular*}{\hsize}{@{\extracolsep{\fill}} lccccccccccccc }
\hline 
\hline 
Cases & $D_{w}$ & $\rho$ & $T$ & $ic_{x}$ & $ic_{y}$ & $ic_{z}$ & $\sigma$ & $b$ &$u_{c}^{\textit{T1Gd}}$ & $u_c^{\textit{FLAIR}}$ & $\sigma^{2}_{\alpha}$ 
\cr 
\hline 
\hline 
P1:  MAP   &  0.188 &  0.029 &  273.130 &  169.293 &  177.075 &  141.235 &  0.111 &  0.959 &  0.693 &  0.521 &  0.051 &  \cr 
P1:  mean  &  0.334 &  0.054 &  192.649 &  169.242 &  177.101 &  140.851 &  0.109 &  0.951 &  0.712 &  0.516 &  0.053 &  \cr 
P1:  std   &  0.190 &  0.031 &  107.396 &  0.435 &  0.384 &  0.435 &  0.013 &  0.040 &  0.029 &  0.016 &  0.002 &  \cr 
\hline 
P2:  MAP   &  0.060 &  0.024 &  337.582 &  108.186 &  158.182 &  173.926 &  0.160 &  0.846 &  0.559 &  0.436 &  0.051 &  \cr 
P2:  mean  &  0.090 &  0.036 &  407.101 &  108.262 &  157.978 &  173.850 &  0.161 &  0.859 &  0.577 &  0.448 &  0.053 &  \cr 
P2:  std   &  0.086 &  0.034 &  257.528 &  0.384 &  0.358 &  0.512 &  0.014 &  0.055 &  0.027 &  0.016 &  0.003 &  \cr 
\hline 
P3:  MAP   &  0.846 &  0.010 &  1094.225 &  112.998 &  155.520 &  117.632 &  0.218 &  0.925 &  0.614 &  0.347 &  0.056 &  \cr 
P3:  mean  &  0.667 &  0.008 &  1468.721 &  111.821 &  154.035 &  118.298 &  0.214 &  0.884 &  0.627 &  0.352 &  0.061 &  \cr 
P3:  std   &  0.133 &  0.002 &  299.056 &  2.611 &  3.046 &  2.074 &  0.004 &  0.029 &  0.011 &  0.013 &  0.003 &  \cr 
\hline 
P4:  MAP   &  0.223 &  0.010 &  848.498 &  163.405 &  162.944 &  183.347 &  0.149 &  1.011 &  0.796 &  0.416 &  0.050 &  \cr 
P4:  mean  &  0.227 &  0.010 &  1015.395 &  163.533 &  162.918 &  183.245 &  0.142 &  1.009 &  0.793 &  0.417 &  0.051 &  \cr 
P4:  std   &  0.167 &  0.007 &  403.933 &  0.333 &  0.410 &  0.307 &  0.010 &  0.009 &  0.005 &  0.007 &  0.000 &  \cr 
\hline 
P5:  MAP   &  2.447 &  0.107 &  93.278 &  117.453 &  110.541 &  141.517 &  0.219 &  1.020 &  0.601 &  0.374 &  0.051 &  \cr 
P5:  mean  &  2.285 &  0.101 &  101.387 &  117.376 &  110.310 &  141.312 &  0.205 &  1.005 &  0.609 &  0.368 &  0.051 &  \cr 
P5:  std   &  0.312 &  0.014 &  17.692 &  0.691 &  0.486 &  0.333 &  0.008 &  0.013 &  0.006 &  0.008 &  0.001 &  \cr 
\hline 
P6:  MAP   &  0.440 &  0.029 &  406.237 &  127.053 &  167.450 &  167.731 &  0.193 &  0.937 &  0.614 &  0.369 &  0.052 &  \cr 
P6:  mean  &  0.869 &  0.053 &  311.425 &  124.442 &  166.093 &  166.912 &  0.190 &  0.949 &  0.642 &  0.379 &  0.053 &  \cr 
P6:  std   &  0.557 &  0.033 &  262.315 &  1.024 &  0.947 &  0.870 &  0.014 &  0.040 &  0.021 &  0.010 &  0.002 &  \cr 
\hline 
P7:  MAP   &  0.196 &  0.007 &  1580.353 &  162.867 &  98.509 &  121.088 &  0.218 &  0.837 &  0.604 &  0.362 &  0.052 &  \cr 
P7:  mean  &  0.282 &  0.010 &  1204.162 &  163.354 &  96.640 &  123.110 &  0.213 &  0.807 &  0.605 &  0.359 &  0.052 &  \cr 
P7:  std   &  0.079 &  0.003 &  340.776 &  0.486 &  1.408 &  1.024 &  0.004 &  0.020 &  0.004 &  0.009 &  0.001 &  \cr 
\hline 
P8:  MAP   &  0.484 &  0.025 &  391.577 &  166.528 &  119.270 &  160.154 &  0.131 &  0.866 &  0.602 &  0.355 &  0.050 &  \cr 
P8:  mean  &  0.689 &  0.034 &  353.465 &  166.093 &  119.219 &  160.410 &  0.130 &  0.845 &  0.605 &  0.355 &  0.051 &  \cr 
P8:  std   &  0.399 &  0.020 &  190.028 &  0.256 &  0.205 &  0.256 &  0.003 &  0.017 &  0.003 &  0.005 &  0.001 &  \cr 
\hline 
\hline 
\end{tabular*} 

\label{table:PatientResults}
\end{table*} 
%%%================================================

\subsection{Effect of the brain anatomy on the tumor cell density}
\begin{figure*}[h]%[tbhp]
\centering
\includegraphics[width=1\linewidth]{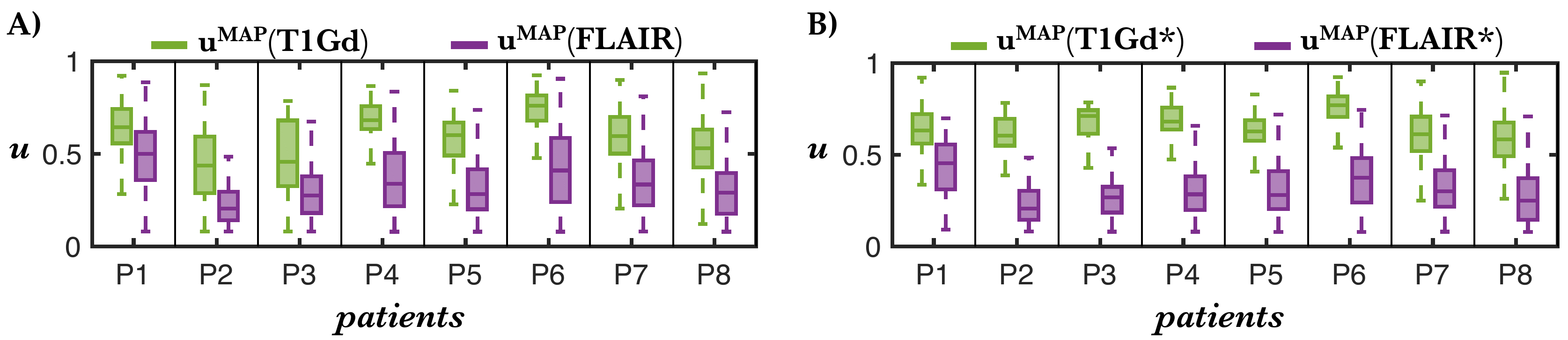}
\caption{The effect of the brain anatomy on the tumor cell density. \textit{A)} Distribution of the inferred tumor cell densities at the border of the T1Gd (green) and FLAIR (purple) tumor segmentations represented by box plots. \textit{B)} Distribution of the tumor cell densities at segmentation borders far from anatomical restrictions. Boxplot whiskers demarcate the $95\%$ percentiles. }
\label{fig:bc}
\end{figure*}

The patient-specific anatomical restrictions and anisotropic tumor growth result in tumor cell densities that are  not constant across the borders of the T1Gd and FLAIR tumor segmentations. Figure \ref{fig:bc} \textit{(A)} shows the distribution of the inferred tumor cell densities across the tumor borders visible in T1Gd and FLAIR scans. The distributions vary significantly along the borders and among the patients. This is not surprising, since in the regions of anatomical restrictions, the outlines of the T1Gd and FLAIR tumor segmentations coincide. The effect of the anatomical restrictions can be alleviated by considering the tumor cell density only at the tumor segmentation outlines far from the ventricles and skull. In this case the inferred tumor cell densities vary less and are also more consistent among the patients as shown in Fig. \ref{fig:bc} \textit{(B)}. This implies the importance of considering the imaging parameters patient-specific, especially when dealing with complex anatomies such as the brain.

%%~~~~~~~~~~~~~~~~~~~~~~~~~~~~~~~~~~~~~~~~~~~~~~~~~~ 
%~~~~~~~~~~~~~~~~~~~~~~~~~~~~~~~~~~~~~~~~~~~~~~~~~~

\subsection{Acquisition protocol}
The MRI scans were conducted with 3T mMR Biograph scanner (Siemens 
Medical Solutions, Germany). FET-PET scans were acquired on a SIEMENS Biograph 64 PET/CT scanner. For standardized metabolic conditions, patients were
asked to fast for a minimum of 6h before the PET scan.  Intravenous administration of 190 MBq of FET was performed, and low-dose CT
(24–26 mAs, 120 kV) for attenuation correction was acquired. Ten-minute PET acquisitions were performed 30 to 40 min post-injection. PET data were reconstructed by filtered back-projection using a Hann filter with corrections for attenuation, scatter, and radioactive decay. All patients provided informed written consent. The study was approved by the ethics committee of the TU Munich and carried out in accordance with the approved guidelines.

\REVa{
\subsection{Image processing}
The scans of each patient were registered to preoperative T1Gd scans using affine ANTs registration \cite{Avants:2009}. The white matter, gray matter, and cerebrospinal fluid were segmented from T1 MRI scan using an atlas-based segmentation approach\footnote{Software for brain tissue segmentation: \url{https://github.com/JanaLipkova/s3}}. In this way, the brain atlas is deformed to reach a spatial consistency with the visible brain tissue of the patient, while the tissue underneath the tumor is described by the anatomy of the deformed atlas. The tumors were segmented with a generative Expectation-Maximization algorithm \cite{Menze:2016}. All segmentations were manually inspected and corrected if needed.
}

\bibliographystyle{IEEEtran}
\bibliography{BibTex/GliomaBibFile.bib}